# RETHINKING GENERALIZED BCIs: BENCHMARKING 340,000+ UNIQUE ALGORITHMIC CONFIGURATIONS FOR EEG MENTAL COMMAND DECODING

**PAUL BARBASTE** [1, 2, 3], **OLIVIER OULLIER** [4, 1, 5] & **XAVIER VASQUES** [6,7]

[1] *Inclusive Brains,* Marseille, France
[2] *Wavestone,* Paris, France
[3] *Human Technology Foundation,* Paris, France
[4] *Human-Computer Interaction Department, Mohamed bin Zayed University of Artificial Intelligence,* Abu Dhabi, UAE
[5] *Institute for Artificial Intelligence,* Biotech Dental Group, Salon-de-Provence, France
[6] *IBM Technology,* Bois-Colombes, France
[7] *IBM France Lab, Orsay,* France

**ASTRACT**

Robust decoding and classification of brain patterns measured with electroencephalography (EEG) remains a major challenge for real-world (i.e. outside scientific lab and medical facilities) brain-computer interface (BCI) applications due to well documented inter- and intra-participant variability. Here, we present a large-scale benchmark evaluating over 340,000+ unique combinations of spatial and nonlinear EEG classification. Our methodological pipeline consists in combinations of Common Spatial Patterns (CSP), Riemannian geometry, functional connectivity, and fractal- or entropy-based features across three open-access EEG datasets. Unlike prior studies, our analysis operates at the per-participant level and across multiple frequency bands (8-15 Hz and 8-30 Hz), enabling direct assessment of both group-level performance and individual variability. Covariance tangent space projection (cov-tgsp) and CSP consistently achieved the highest average classification accuracies. However, their effectiveness was strongly dataset-dependent, and marked participant-level differences persisted, particularly in the most heterogeneous of the datasets. Importantly, nonlinear methods outperformed spatial approaches for specific individuals, underscoring the need for personalized pipeline selection. Our findings highlight that no universal "one-size-fits-all" method can optimally decode EEG motor imagery patterns across all users or datasets. Future work will require adaptive, multimodal, and possibly novel approaches to fully address neurophysiological variability in practical BCI applications where the system can automatically adapt to what makes each user unique.

**Correspondence:**
Paul Barbaste      paul@inclusive-brains.com
Olivier Oullier      olivier.oullier@mbzuai.ac.ae
Xavier Vasques      xaviervasques@fr.ibm.com

**INTRODUCTION**

In recent years, brain-computer interfaces (BCIs) have garnered significant attention as invasive[1] and non-invasive[2] tools for enabling "direct" communication between the brain and external connected devices and digital environment[3] both in the scientific and medical communities as well as in mainstream media. BCIs are multidisciplinary systems enabling individuals with severe motor disabilities such as locked-in syndrome, amyotrophic lateral sclerosis, spinal cord injury, or severe movement or speech disorders following stroke[4–8] to be able to communicate and sometimes, even move again even when the aforementioned conditions prevented them from doing it.

BCI systems rely on interpreting neural activity to "decode" mental states and intentions to translate them into commands, often referred to as "mental commands", empowering their users equipped with intra-cranial or surface electrodes to control connected devices such as wheelchairs, prosthetics, computers, workstations or vocal synthesis[5,9–11].

The most popular neurotechnology for non-invasive BCIs is EEG[2]. However, despite remarkable progress, decoding cognitive states such as motor imagery: the mental simulation of movement that has been reported to exhibit overlapping brain dynamics with the execution of movements[12] from EEG data remains challenging because of the intrinsic variability of brain signals[13] and the noise typical of non-invasive recordings[5,14,15].

EEG provides high temporal resolution but is notoriously susceptible to both inter- and intra-session variability[16]. EEG signal can fluctuate markedly between individuals and across sessions recorded on the same individual due to factors such as attention, fatigue, electrode shifts, impedance changes, and environmental conditions[3]. This non-stationarity complicates the development of generalized models that would accurately decode mental states across diverse participants and experimental settings, something that is a critical requirement for the use of BCIs in medical and everyday life settings[5]. The phenomenon commonly termed "BCI illiteracy,"[10,17,18] is reported to affect 15-30% of users who are unable to achieve adequate control, underscores the need for more generalizable and adaptive decoding approaches[19,20].

Machine Learning (ML) and Deep learning (DL) techniques have shown promise in addressing some of these challenges[21–23]. Classical ML algorithms such as support vector machines (SVMs)[24] and linear discriminant analysis (LDA)[25], as well as convolutional neural networks (CNNs)[26], are widely used for EEG classification[5,27]. In addition, spatial feature extraction has long been central to EEG decoding. Two prototypical approaches are Common Spatial Patterns (CSP)[28] and Riemannian geometry-based methods that operate on trial covariance matrices (with tangent-space projections for Euclidean classifiers)[29–31]. However, due to significant non-stationarity and inter-subject variability in EEG, exclusively spatial methods often fail to generalize across users, resulting in performance drops when applied to new participants or varying recording conditions[32].

Consequently, benchmarking a broader spectrum of EEG-derived features has become essential to improving the performance and accuracy of non-invasive BCIs. Feature extraction techniques are highly diverse and play a critical role in EEG analysis approaches range from one-dimensional methods (such as time-domain, frequency-domain, and decomposition-domain features) to multi-dimensional techniques, highlighting their importance across numerous applications[33]. Similarly, transforming raw EEG signals into meaningful representations, including time-domain statistics, frequency-domain features (e.g., power spectral density), and time-frequency methods (e.g., wavelet transform), is crucial for accurate signal classification and robust performance[34,35]. One should also underline that cross-user generalization depends on diverse and transferable features, noting that non-stationarity and dataset shift in EEG-based emotion recognition requires transfer learning and domain generalization strategies built upon robust feature representations[36].

In parallel with spatial techniques (e.g., CSP and Riemannian tangent-space projections), nonlinear descriptors seek to capture the complex dynamics of EEG and may better reflect intra-subject variability related to fatigue or fluctuating attention. Common examples include Hjorth parameters, Higuchi's fractal dimension, Hurst exponent, Fisher information-based measures, and singular value decomposition-based entropy (SvEn)[14,37–39]. Beyond univariate features, functional-connectivity measures provide complementary information about task-related network interactions in EEG and are increasingly considered in patterns of brain activity underlying motor-imagery pipelines[40–42].

An open and critical question remains: *Which combinations of feature-extraction and classification techniques yield truly robust and generalizable performance, especially in the presence of substantial individual variability?* Answering this requires evaluation protocols that minimize optimistic bias for starters. In practice, this means using participant-independent validation strategies (e.g., leave-one-subject-out (LOSO) cross-validation)[43], employing performance metrics robust to class imbalances (such as balanced accuracy and AUROC)[44], assessing statistical significance via nonparametric or permutation testing, and rigorously guarding against data leakage. Particularly by ensuring that operations like filter estimation and feature normalization are confined strictly to the training folds and never informed by the test data[45,46].

Here, we present what is, to our knowledge, the most extensive and systematic benchmark of spatial and non-linear EEG decoding pipelines for motor imagery (MI) to date, conducted in the context of brain-computer interface (BCI) applications. We evaluated 342,604 combinations of feature extraction and classification methods, including spatial filters[47] such as common spatial patterns (CSP)[28], Riemannian geometry based analyses[48], tangent-space projections of covariance matrices[49], functional connectivity[50–52] and a broad spectrum of non-linear descriptors[53], including Hjorth parameters[15], singular value decomposition (SVD) entropy[54], Hurst exponent[37], Fisher information[55] and Higuchi fractal dimension (HFD)[56].

The benchmark was conducted across three publicly available EEG datasets widely used in BCI research:
- The PhysioNet EEG Motor Movement/Imagery dataset (109 participants, 64 electrodes)[57];
- The Zhou2016 dataset (4 participants, 14 electrodes)[39];
- The Cho2017 dataset (52 participants, 64 electrodes, including concurrent EMG recordings and questionnaire data)[14].

To ensure reproducibility and facilitate comparison with future studies, we align our preprocessing and validation procedures with established community standards, most notably those introduced by the *Mother of All BCI Benchmarks* (MOABB) initiative[58,59].

Distinct from prior work that primarily reports aggregated group-level performance[60], our research adopts a per-subject analytical framework, enabling fine-grained assessment of inter-individual variability. We quantify both average performance and inter-individual variability and believe this work yields a detailed mapping of *"what (does not) works, for whom, and under which conditions"*, i.e. which specific pipelines might perform better than others. Our goal is to provide a rigorous scientific approach that leads to practical insights into the personalization and clinical translation of EEG decoding systems that will improve the performance of real-world BCI systems using this neurotechnology.

**MATERIALS & METHODS**

**Datasets**

In order to evaluate the replicability of our approach, we used three different open-source datasets, using the Mother Of All BCI Benchmark (MOABB) framework, introduced by Jayaram and colleagues[59,61]. In this study, we incorporated the Physionet Motor Imagery (MI) dataset into the MOABB framework. This dataset includes over 1,500 electroencephalography (EEG) recordings. Each recording, lasting between one and two minutes, was collected from 109 volunteers using the BCI2000 system. This integration was facilitated by referring to the documentation available on BCI2000. The recordings were captured through a 64-channel device operating at a sampling frequency of 160 Hz[57]. Each participant performed fourteen series of experiments comprising two one-minute baseline series (one with eyes open and one with eyes closed), as well as three two-minute series for each of the following four motor/imagined tasks:

- Appearance of a target on the left or right side of the screen, leading the subject to open and close the corresponding fist until the target disappeared, followed by a period of relaxation.

- Appearance of a target on the left or right side of the screen, prompting the subject to imagine opening and closing the corresponding fist until the target disappears, followed by a period of relaxation.

- Appearance of a target at the top or bottom of the screen, prompting the subject to open and close both fists (if the target is at the top) or both feet (if the target is at the bottom) until the target disappears, followed by relaxation.

- Appearance of a target at the top or bottom of the screen, prompting the subject to imagine opening and closing both fists (if the target is at the top) or both feet (if the target is at the bottom) until the target disappears, followed by relaxation.

We also used the Zhou2016[39]. This dataset was collected from 4 subjects performing three types of motor imagery: left hand, right hand and feet. Each subject participated in three data collection sessions. Each of these sessions consisted of two consecutive series interspersed with pauses of several minutes, with each series including 75 trials divided equally between the three classes (i.e. 25 trials per class). The intervals between sessions varied from several days to several months, providing significant temporal diversity in data collection. Trials began with a short beep indicating a one-second preparation time, followed by the display of a red arrow pointing randomly in one of three directions (left, right or down) for 5 seconds. This was then followed by a black screen for 4 seconds. Subjects were instructed to immediately perform the imagination tasks corresponding to the movement of the left hand, right hand or feet according to the direction of the arrow, and to attempt to relax while the black screen was displayed. Data were recorded using a 14-channel EEG device with a sampling frequency of 250 Hz. Each class of imagined movement was subjected to 160 trials, spread over the three sessions, for a total of 480 trials per subject, each lasting 5 seconds.

Lastly, we used the dataset provided by Cho and al.[14], using MOABB. This dataset was collected with the aim of understanding the motor imagery (MI) dynamics of left- and right-hand movements. The experiment involved 52 subjects (19 females, with a mean age and standard deviation of 24.8 ± 3.86 years). Among them, subjects s20 and s33 were ambidextrous, while the other 50 were right-handed. The authors used 64 active Ag/AgCl electrodes for EEG data recording, with a set-up based on the international 10-10 system. EEG signals were collected at a sampling rate of 512 Hz using the Biosemi ActiveTwo system. In addition, the BCI2000 version 3.0.2 system was used to collect EEG data and present motor imagery (MI) movement instructions for the left and right hands. During the experiment, each subject was asked to imagine the hand movement corresponding to the instruction given. Five to six series were performed during the MI experiment, after which the classification accuracy for each series was calculated and communicated to the subjects in order to boost their motivation. A break of up to 4 minutes was allowed between each run, depending on the subject's needs.

In addition to using the MOABB framework and the datasets described, we applied a specialized configuration using the LeftRightImagery paradigm to capture motor imagery tasks from the EEG signals. This paradigm focuses on classifying left- and right-hand motor imagery by extracting features from specific channels. We choose two frequency bands, respectively 8-15 and 8-30 Hz, a time window of 0.6-2 seconds and two classes corresponding to left and right-hand motor imagery. This setup was used consistently across the three datasets, with slight adjustments in the sampling frequencies according to the dataset specifications such as the sampling frequency set to 250 Hz for Zhou2016, 160 Hz for PhysionetMI and 512 Hz for Cho2017.

By focusing on specific channels and optimizing the feature extraction accordingly, we ensured that the data captured relevant motor and premotor brain activities for effective classification of the motor imagery tasks.

**EEG Signal Processing**

**Filtering**

In our preliminary experiments, we explored a broad spectrum of frequencies ranging from 8 to 30 Hz. Subsequently, we refined our methodology by implementing specific bandpass filtering, concentrating our analysis on sensorimotor rhythms, particularly the mu rhythm which spans 8 to 15Hz. To isolate the signals of interest and mitigate the influence of evoked responses that could bias data interpretation, we devised a targeted temporal analysis strategy. Our approach involved analyzing epochs that commenced one second post-stimulus, thereby excluding the immediate post-stimulus interval that is typically saturated with evoked responses. This shift allowed us to focus on the ensuing spontaneous brain activity. By emphasizing these more stable attributes indicative of the subject's sensorimotor state, our method enhances the accuracy of brain signal classification.

**EEG Feature Extraction**

**Common Spatial Patterns (CSP)**

Different combinations of data preprocessing techniques were applied before training the classical machine learning algorithms. First, we used signal processing methods such as Common Spatial Patterns (csp), widely used in EEG signal classification. In the context of EEG, CSP was first described by Koles et al.[62]. This technique is employed to extract and discriminate features by optimizing variance between two different motor imagery tasks or two different mental states[45,63,64]. This involves the transformation of EEG signals into a new vector space, where the variance of the associated signals, associated with one class (mental state or motor task) is maximized, while that corresponding to the other class is minimized. The goal of this method is to identify and apply spatial filters to highlight significant differences between several classes. CSP is a multi-stage process which begins with the calculation of covariance matrices for each EEG signal class. The covariance between two variables X and Y can be calculated using the following formula:

$$Cov(X,Y) = E[(X - \mu X) \cdot (Y - \mu Y)]$$

where $E$ represents the mathematical expectation, $\mu X$ is the mean of $X$, and $\mu Y$ is the mean of $Y$. For a series of $n$ observations:

$$X = (x_1, x_2, \ldots, x_n)$$
$$Y = (y_1, y_2, \ldots, y_n)$$

The formula therefore becomes:

$$Cov(X,Y) = \frac{1}{n-1} \sum_{i=1}^{n} (x_i - \underline{x})(y_i - \underline{y})$$

where $\underline{x}$ et $\underline{y}$ are means of X and Y respectively.

For EEG signals, where $X$ and $Y$ are representing signals from two different EEG channels, the covariance measures the signals' joint variability. When the covariance between all pairs of channels in an EEG dataset is calculated, this results in a covariance matrix, where each element $C_{ij}$ represents the covariance between channels $i$ and $j$. The covariance matrix for a multichannel EEG is then:

$$C = \frac{1}{n-1} \sum_{i=1}^{n} (X_t - \underline{X})(X_t - \underline{X})^T$$

Where $X_t$ denote the signal vector at time $t$, which encapsulates the signals from all channels at that moment. The vector $\underline{X}$ represents the average for each channel, computed over the entire observation period. The transpose $T$ flips a vector into a row or column format, allowing it to be multiplied by itself in a way that generates a matrix of all possible combinations of its elements. The CSP then uses linear algebra techniques (e.g. the eigenvalue decomposition) to determine the associated eigenvectors (i.e. the spatial directions) that will maximize the signal variance contrast between the examined classes. This reveals discriminating features that are suited to the classification task. For a pair of covariance matrices $A$ and $B$ (representing the covariances of the EEG signals for two different mental states or motor tasks), the goal is to solve the generalized eigenvalue problem:

$$Av = \lambda Bv$$

where $\lambda$ are the eigenvalues and $v$ are the eigenvectors. For CSP, this translates into finding the directions (eigenvectors) that maximize the ratio of the variances between the two states. This is often done by calculating the eigenvalues and eigenvectors of $B^{-1}A$ or equivalently $A^{-1}B$ if this simplifies the calculations, depending on the condition of the matrices. The output of this step are the eigenvectors $v$ which will act as spatial filters, and the eigenvalues $\lambda$ which indicate the significance (or variance) associated with each eigenvector.

Once the eigenvectors $v$ and eigenvalues $\lambda$ have been obtained, the selection of components in CSP consists of choosing the eigenvectors that correspond to the largest (and smallest, depending on the case) eigenvalues, as they represent the directions that maximize (or minimize) the variance contrast between classes. Selection is often made by ordering the eigenvectors according to their corresponding eigenvalues, from largest to smallest. The first $n$ eigenvectors (where $n$ is a user-defined parameter, for example $n_components$ in the code) are selected to transform the original EEG data into a new space where the signals are better separated according to the classes of interest. In practice, this is achieved by projecting the EEG data onto these selected eigenvectors:

$$X_{transformed} = v_{selected}^T X$$

where $X$ is the matrix of the original EEG signals and $v_{selected}$ contains the selected eigenvectors. $X_{transformed}$ then represents the EEG data in the new CSP space, ready for further analysis or a classification task. This approach allows meaningful features to be extracted from the EEG data, focusing on the spatial components that are most discriminative between the mental states or motor tasks under investigation. In this context, it is pertinent to mention the Filter Bank Common Spatial Patterns (FBCSP) method, as introduced by Ang et al.[65]. This method advances the traditional CSP by integrating bandpass filters to decompose the signal into frequency sub-bands prior to CSP application. The FBCSP is engineered to enhance feature discrimination by harnessing information specific to each frequency band. However, in our study, we chose not to utilize FBCSP; instead, we focused on directly applying CSP to analyze EEG signals associated with motor imagery, without engaging in sub-band decomposition.

**Riemannian geometry - Tangent Space Projection (TGSP) on covariance matrices**

Riemannian geometry was also experimented. The Riemannian geometry for the classification of EEG signals has been widely used since the last decade, first introduced by Barachant and colleagues[19,45]. It is particularly relevant for EEG classification because it allows matrices to be manipulated with fluidly defined positivity constraints. Riemannian manifolds (or Riemannian varieties) are complex mathematical structures that allow linearisation via tangent space, thus facilitating the processing of EEG data that is intrinsically non-linear and very high dimensional. Approaches leveraging Riemannian geometry have attained leading-edge success in a range of BCI applications[19]. Tangent Space Projection (TGSP) is a mathematical technique used to process complex data, such as brain signals captured by EEG[45,66]. This technique is specifically applied to the manipulation of symmetric positive definite (SPD) matrices. SPD matrices are very common in neuroscience and signal analysis, as they can represent, for example, the covariance structure of EEG signals, providing valuable information about brain activity. The space of SPD matrices is not flat but curved (or non-Euclidean).

This makes direct analysis of these matrices more complex, as conventional linear operations (such as mean or variance) are not directly applicable in any meaningful way. TGSP allows us to get round this problem by projecting the SPD matrices into a tangent space. This tangent space is a linear (Euclidean) space where conventional operations can be applied more easily. TGSP maps each SPD matrix to a point in a tangent space, generally chosen at an average point of the set of matrices. This projection transforms the matrices into vectors of a Euclidean space, enabling standard analysis and classification techniques to be used. The dimension of the tangent space is determined by the size of the SPD matrices, with each vector in the tangent space having a dimension of $\frac{n(n+1)}{2}$ where n is the dimension of the SPD matrix. The aim is to project these matrices into a tangent space where they can be manipulated like vectors in a Euclidean space, and then potentially perform the inverse operation to find the SPD matrices. The projection of an SPD matrix $X_i$ on the tangent space at the reference point $C_r$ (the geometric mean covariance matrix of the set of SPD matrices) is given by:

$$T_i = \Phi_{C_r}(X_i) = \log_{C_r}(X_i)$$

Where $T_i$ is the projection of $X_i$ on the tangent space at the point $C_r$, $\Phi_{C_r}$ is the projection application, $\log_{C_r}$ is the generalized matrix logarithm that maps $X_i$ from the Riemannian variety to the tangent space. This operation is performed for each SPD matrix in the set $X$.

The result of this projection is a set of "flattened" matrices into vectors of dimension $\frac{n(n+1)}{2}$, where $n$ is the number of channels (or the dimension of the SPD matrices). The reference point $C_r$ is crucial for these operations. It is usually calculated as the geometric mean of the set of SPD matrices X:

$$C_r = \text{mean}_{geom}(X)$$

This geometric mean serves as the optimal "center" for the projection onto the tangent space, minimizing the local distortion of geometric information inherent in the variety of SPD matrices.

**Riemannian geometry - Tangent Space Projection (TGSP) on coherence matrices (functional connectivity)**

In this study, we used instantaneous coherence to characterize functional connectivity between different brain regions based on EEG signals. The interest of using instantaneous coherence to classify EEG signals lies in its ability to capture in-phase functional interactions between brain regions, providing robust features for differentiating cognitive states or experimental conditions. We chose to focus on instantaneous coherence in order to specifically capture phase synchronization between brain regions. Unlike other types of coherence, this measure offers a direct assessment of synchronous neuronal interactions, essential for the classification of cognitive states and motor imagery. For each pair of channels, instantaneous coherence is calculated using the following formula:

$$C(f) = \frac{Re\left(S_{xy}(f)\right)^2}{S_{xx}(f) \cdot S_{yy}(f)}$$

Where:

$S_{xy}(f)$: the cross-spectrum between signals $x$ and $y$ at frequency $f$,
$S_{xx}(f)$ and $S_{yy}(f)$:: the respective power spectra of signals $x$ and $y$,
$Re\left(S_{xy}(f)\right)$: the real part of the cross-spectrum $S_{xy}(f)$.

This measure provides functional connectivity matrices for each temporal segment of the EEG signals. These instantaneous coherence matrices, which are symmetric positive definite (SPD) matrices, contain information about the strength of neuronal interactions at different frequencies. To exploit these matrices for classification purposes, we applied Tangent Space Projection (TSP), previously explained.

**Non-Linear Features**

In this study, we also evaluated non-linear feature extraction methods to capture the inherently irregular and fractal nature of EEG signals. Unlike traditional linear approaches, these non-linear transformations provide insights into the complex neuronal interactions essential for classifying cognitive and motor states. The strength of non-linear methods lies in their ability to capture intricate dynamics, offering robust features for distinguishing between various cognitive states and experimental conditions. By analyzing the complexity and variability of EEG signals, non-linear features reveal patterns that may remain undetected by conventional methods.

**Higuchi Fractal Dimension**

The **Higuchi Fractal Dimension (HFD)** is computed by analyzing the signal at multiple scales $k$. For each scale, the signal is segmented into k distinct subsequences, and the length $L_{m,k}$ of each subsequence is computed as:

$$L_{m,k} = \frac{(N-1)}{\left\lfloor \frac{N-m}{k} \right\rfloor \cdot k} \sum_{i=1}^{\left\lfloor \frac{N-m}{k} \right\rfloor} |X[m + i \cdot k] - X[m + (i-1) \cdot k]|$$

The average length $L(k)$ is then calculated for each $k$, and a log-log linear regression between $log(L(k))$ and $log(1/k)$ is performed. The **slope** of this regression provides the Higuchi Fractal Dimension, offering a precise estimate of the signal's complexity. In this study, we set the number of scales $k$ to 10, meaning that the HFD was computed using 10 subdivisions (segments) of the signal. This parameter determines the granularity of the fractal dimension estimation and directly influences the sensitivity of the HFD to the signal's complexity.

**Hjorth Parameters**

Hjorth Parameters[15] are widely used to describe the dynamic characteristics of EEG signals. They consist of two key metrics which are the mobility, representing the rate of change in the signal, providing a measure of its smoothness and the complexity quantifying the deviation of the signal from a pure sine wave, reflecting the intricacy of the brain activity. These parameters are particularly useful for characterizing both the stability and complexity of neural activity.

To characterize the dynamic properties of the EEG signals, we computed the Hjorth parameters (mobility and complexity) directly from each time series. Let $X$ denote the signal of length $n$, and $D$ its first-order difference sequence, where $D[i] = X[i] - X[i-1]$ with $D[0] = X[0]$.

The total power of the signal was calculated as:

$$TP = \sum_{i=1}^{n} X[i]^2$$

The mean squared value of the first difference was computed as:

$$M2 = \frac{1}{n} \sum_{i=1}^{n} D[i]^2$$

The mean squared value of the second difference was computed as:

$$M4 = \frac{1}{n} \sum_{i=2}^{n} (D[i] - D[i-1])^2$$

Hjorth mobility was then defined as:

$$\text{Mobility} = \sqrt{\frac{M2}{TP}}$$

and Hjorth complexity as:

$$\text{Complexity} = \sqrt{\frac{M4 \cdot TP}{M2^2}}$$

**Singular Value Decomposition Entropy**

Singular Value Decomposition (SVD) entropy measures the complexity of EEG signals by performing a singular value decomposition on the embedding matrix constructed from the time series. This method identifies the dominant modes within the signal and provides a quantitative summary of its complexity, revealing underlying neural patterns not immediately apparent through linear analysis.

We computed the SVD entropy as follows. For a given time series $X$, we first constructed an embedding matrix $M$ using time delay $\tau$ and embedding dimension $dE$:

$$M = \begin{bmatrix} X[0] & X[\tau] & \cdots & X[(dE-1)\tau] \\ X[1] & X[1+\tau] & \cdots & X[1+(dE-1)\tau] \\ \vdots & \vdots & \cdots & \vdots \end{bmatrix}$$

We then performed singular value decomposition on $M$ to obtain the singular values $Wi$, which were normalized as:

$$w_i = \frac{W_i}{\sum_j W_j}$$

Finally, SVD entropy was calculated as:

$$H_{\text{SVD}} = -\sum_i w_i \log w_i$$

This entropy value reflects the diversity of the singular value spectrum and thus the structural complexity of the signal.

**Hurst Exponent**

The Hurst Exponent assesses the long-term memory of a time series, indicating whether a signal behaves randomly (H ≈ 0.5), follows a persistent trend (H > 0.5), or displays anti-persistent behavior (H < 0.5). For EEG signals, this measure is useful in detecting sustained patterns or unpredictability in neural dynamics.

We computed the Hurst exponent as follows. Let $X$ be a time series of length $N$. We first calculated the cumulative sum process $Y$, where:

$$Y_t = \sum_{i=1}^{t} X[i]$$

for $t = 1, 2, \ldots, N$. The mean up to each time point was then

$$\overline{Y}_t = \frac{Y_t}{t}$$

For each time $t$, we computed the standard deviation of the signal up to $t$,

$$S_t = \text{std}(X[1:t])$$

and the range of the adjusted cumulative sum,

$$R_t = \max(Y[1:t] - t \cdot \overline{Y}_t) - \min(Y[1:t] - t \cdot \overline{Y}_t)$$

We then calculated the rescaled range for each $t$:

$$\frac{R_t}{S_t}$$

Taking the logarithm of the rescaled range and the logarithm of the time index yields two sequences. The Hurst exponent $H$ was estimated as the slope of the least-squares linear regression between $log(Rt/St)$ and $log(t)$:

$$\log\left(\frac{R_t}{S_t}\right) = H \cdot \log(t) + c$$

**Data preprocessing**

Before implementing machine learning algorithms on the selected datasets, initial investigations assessed the accuracy of the classification models and the impact of different feature engineering methods. For this purpose, we conducted a series of experiments that included feature rescaling, extraction, and selection, using classical classification algorithms. Further analyses focused on applying the following feature rescaling techniques:

StandardScaler: $z = \frac{x - mean(x)}{standard\ deviation\ (x)}$

MinMaxScaler: $z = \frac{x_i - (x)}{(x) - (x)}$

MaxAbsScaler: $z = \frac{x_i}{(abs(x))}$

RobustScaler: $z = \frac{x_i - Q_1(x)}{Q_3(x) - Q_1(x)}$

l2-normalization: y= $||x||_2 = \sqrt{\sum_{i=1}^{n} x_i^2}$

Logistic data transformation: $z = \frac{1}{1 + e^{-x}}$

Lognormal transformation: $F(x) = \phi\left(\frac{ln ln\ (x)}{\sigma}\right), x \geq 0\ ;\ \sigma \geq 0$

Yeo-Johnson:

$$\psi(\lambda, y) = \begin{cases} \frac{(y+1)^\lambda - 1}{\lambda} & if\ \lambda \neq 0, y \geq 0 \\ \log(y+1) & if\ \lambda = 0, y \geq 0 \\ -\frac{[(-y+1)^{(2-\lambda)} - 1]}{2 - \lambda} & if\ \lambda \neq 2, y < 0 \\ -\log(-y+1) & if\ \lambda = 2, y < 0 \end{cases}$$

The nonparametric quantile transformation, which includes both normal and uniform distributions, adjusts the dataset to conform to a specified probability distribution, such as normal or uniform. This adjustment is accomplished by using the quantile function, which is essentially the inverse of the cumulative distribution function, on the dataset. Consider a random variable X that follows a normal distribution:

$$X \sim N(\mu, \sigma^2)$$

Then, the quantile function of X is

$$Q_X(p) = \sqrt{2\sigma} \cdot erf^{-1}(2p - 1) + \mu$$

where erf$^{-1}$(x) is the inverse error function.

Let X be a random variable following a continuous uniform distribution:

$$X \sim U(a, b)$$

Then, the quantile function of X is

$$Q_X(p) = \{-\infty, \quad if\ p = 0 \ \ bp + a(1-p), \quad if\ p > 0$$

**Classical Machine Learning**

In this study, several classical machine learning classifiers were employed to analyze EEG data, focusing on classifying motor imagery tasks. The classification process involved applying a combination of feature extraction methods and preprocessing techniques, followed by the use of different classification algorithms. We applied logistic regression with elastic-net regularization (a combination of L1 and L2 penalties). Elastic-net regularization is effective for handling sparse data and preventing overfitting by balancing the L1 and L2 penalties. The solver used for optimization was the saga solver, which is well-suited for large datasets. The intercept scaling was set to 1000 to handle features on different scales, and a random state of 42 was applied for reproducibility. A random forest classifier was also implemented to classify EEG signals based on an ensemble of decision trees. Random forest builds multiple decision trees and aggregates their predictions to improve accuracy and reduce overfitting. It is particularly effective at handling non-linear interactions between features. This algorithm was less prone to overfitting compared to individual decision trees. Both linear support vector machine (SVM) and radial basis function (RBF) kernel SVM were utilized for classification.

The linear SVM was used to separate linearly separable data, such as features from CSP. RBF Kernel SVM, on the other hand, was applied to non-linearly separable data by mapping the features into a higher-dimensional space to capture complex relationships. Linear Discriminant Analysis (LDA) was used as an additional classification method. LDA maximizes the ratio of between-class variance to within-class variance in the data, making it a powerful tool for linearly separable datasets. We explored a wide range of multilayer perceptrons (MLPs) with different architectures to classify EEG data. The MLP configurations included different numbers of hidden layers and neurons. Specific architectures included basic MLP architectures with a single hidden layer, more complex models with three hidden layers, such as configurations of (10, 30, 10) or (20,) neurons, using ReLU activation functions, MLPs with logistic activation functions, trained using stochastic gradient descent (SGD) solvers with constant learning rates or MLP variants with adam solvers, which were tuned to manage adaptive learning rates. MLPs incorporated L2 regularization through the alpha parameter to prevent overfitting. Different values of alpha were tested to control the amount of regularization applied to the model. A constant learning rate was used in several MLP configurations to ensure steady convergence during training. Both SGD and adam solvers were fine-tuned to optimize performance across various feature extraction methods. For reproducibility and traceability, each unique architecture was indexed (MLP_1-MLP_17) and used consistently across all pipeline combinations.

**Assessment**

The performance of the classification algorithms was evaluated using a within-session evaluation approach, as described by Jayaram and Barachant[59] in their work on trustworthy algorithm benchmarking for brain-computer interfaces (BCIs). The MOABB framework was employed to facilitate consistent benchmarking across different EEG datasets, ensuring a robust evaluation of the machine learning algorithms used to classify motor imagery tasks.

In this setting, each subject's recording session is partitioned into predefined training and testing splits defined by MOABB, ensuring that every classifier is evaluated on identical data partitions. This evaluation paradigm isolates the model's ability to generalize to unseen trials *within the same recording session*, a setting that is representative of practical closed-loop BCI use where the dominant sources of variability originate from transient fluctuations in cognitive state, attention, or sensorimotor rhythms rather than long-term session-to-session drift.

For each dataset and frequency band, every pipeline was trained on the session-specific training subset and evaluated on the corresponding held-out test subset, yielding one accuracy score per subject. These subject-wise accuracies were then aggregated to compute the mean performance and dispersion of each pipeline. Unlike cross-validation procedures that artificially resample data, the MOABB within-session protocol enforces a single, deterministic split per session, thereby preventing information leakage and enabling fair comparison across hundreds of pipelines and feature representations. This strict standardization is essential for large-scale benchmarking: it ensures that differences in performance can be attributed to the feature extraction method, scaling strategy, and classifier architecture rather than to idiosyncrasies in data partitioning.

All evaluations were conducted independently for each dataset (PhysionetMI, Zhou2016, and Cho2017), each frequency band (8-15 Hz and 8-30 Hz), and each CSP dimensionality or feature configuration. To ensure reproducibility, the random state of all classifiers employing stochastic optimization was fixed to 42. The resulting performance distributions provide a comprehensive quantification of how classical machine learning models behave under controlled, session-specific conditions that closely approximate real-time BCI operation.

**Within-Session Evaluation**. Implemented using the following key parameters:

**Paradigm**. The specific motor imagery task being evaluated (e.g., left-hand vs. right-hand motor imagery).

**Datasets**. Each dataset (e.g., PhysionetMI, Zhou2016, Cho2017) was treated individually for within-session evaluation, allowing us to assess performance across multiple open-source EEG datasets.

**Feature Dimension.** The number of features used for classification was adjusted depending on the feature extraction method, such as CSP or Cov+TGSP.

**Frequency Bands.** We evaluated the models over different frequency bands (typically 8-15 Hz and 8-30 Hz) to assess how different frequency ranges impact classification accuracy.

**Random State**. A fixed random state of 42 was used to ensure reproducibility across multiple runs of the algorithm.

The evaluation function generated accuracy scores for each classifier, which were averaged to assess the overall performance of each machine learning pipeline within the session. This method ensured that the models were tested in realistic conditions, with data variability captured from a single recording session, closely mimicking the real-world use of BCIs in practice.

**Code availability**

The algorithms utilized in this study were implemented using Python 3 [Python's official website] (http://www.python.org). For classical computations and manipulations, we employed the widely used open-source library, [scikit-learn] (https://scikit-learn.org/stable/). We utilized PyRiemann, a Python library designed for handling covariance matrices derived from multivariate time series data. PyRiemann is particularly adept at providing tools for covariance and coherence matrix manipulation and advanced algorithms for classification and feature extraction based on Riemannian geometry. This library facilitated the implementation of geometric methods to compute the means and distances of covariance matrices, which are crucial for the effective classification of brain-computer interface (BCI) signals. We used PyEEG to implement the nonlinear methods.

We employed MNE-Python, an open-source Python library developed specifically for the analysis of neurophysiological data, including magnetoencephalography (MEG) and electroencephalography (EEG). MNE-Python supports a wide range of data processing capabilities from raw data preprocessing to advanced statistical analysis and visualization. Additionally, we integrated MOABB (Mother of All BCI Benchmarks) into our analysis pipeline to load datasets and facilitate the evaluation and comparison of different algorithms.

**RESULTS**

In this study, we evaluated a wide range of machine learning pipelines that included feature rescaling, feature reduction, classifiers, and feature extraction techniques applied to motor imagery EEG signals. We began by comparing the covariance projection method in tangent space (cov-tgsp), a state-of-the-art approach[45], with other techniques such as common spatial patterns on covariances (csp)[46], functional connectivity projected onto tangent space (con_instantaneous + tgsp)[50], and non-linear features including Hjorth parameters (hjorth)[15], Higuchi fractal dimension (hfd)[38], and singular value decomposition entropy (svd_entropy)[48]. Additional features such as the Hurst exponent[67], Petrosian fractal dimension[68], Fisher information[55], approximate entropy[53], and detrended fluctuation analysis[69] were also evaluated.

The EEG data was derived from motor or mental imagery tasks using the PhysionetMI[40], Zhou2016[39], and Cho2017[14] datasets. For cov-tgsp and con_instantaneous+tgsp, 2080 features were extracted, 64 for svd_entropy and hfd, 192 for Hjorth and two components for CSP. Tests were performed across two frequency bands, 8-15 Hz and 8-30 Hz, to evaluate the performance of these methods under specific frequency conditions.

We first conducted a general benchmark on PhysionetMI and Zhou2016 (Figure 1), resulting in 243,250 evaluations for 109 subjects in PhysionetMI and 98,730 for 4 subjects in Zhou2016 (see Supplementary Materials 1a, 1b, 1c). In this work, pipeline performance was quantified using the *within-session* evaluation procedure implemented in MOABB, which performs repeated k-fold cross-validation on EEG data originating from a single recording session. This ensures that all preprocessing, feature extraction, and model fitting steps are performed strictly within the training folds, thereby avoiding data leakage and providing a realistic estimate of performance under session-specific variability. For each pipeline and subject, MOABB outputs an accuracy score reflecting the model's capacity to discriminate left- versus right-hand motor imagery. Accuracy is the standard metric reported by the MOABB framework and is used consistently throughout our analyses.

**Overall assessment**

We grouped the data from the general benchmark by pipeline and calculated the average performance score for each method across all subjects and datasets. The methods were then ranked in descending order based on their average scores, highlighting the highest-performing machine learning pipelines (see Supplementary Material 2a, 2b, 2c). Logistic regression consistently yielded the highest mean performance across subjects, particularly when paired with StandardScaler, RobustScaler, or Normalizer and cov_tgsp as the feature extractor for both PhysionetMI and Zhou2016 datasets for both frequency ranges.

Subsequently, we conducted an additional benchmark using logistic regression as the main classifier, along with robust and standard scaling techniques. This evaluation included a diverse set of feature extraction methods: CSP with two components, cov_tgsp, con_instantaneous+tgsp, Hjorth parameters, Higuchi Fractal Dimension (HFD), and singular value decomposition entropy (svd_entropy) (see Supplementary Material 3a). In addition to PhysionetMI and Zhou2016, we incorporated the Cho2017 dataset into this benchmark, resulting in an extra 624 evaluations.

Across all datasets and frequency bands (8-15 Hz and 8-30 Hz), the cov_tgsp pipeline outperformed all others, with a mean score of 0.69 (see Table 1, Figure 2). The CSP pipeline followed closely, scoring 0.66 on the 8-30 Hz band and 0.64 on the 8-15 Hz band. In contrast, pipelines based on con_instantaneous+tgsp, HFD, Hjorth, and svd_entropy showed lower effectiveness, with mean scores ranging from 0.53 to 0.61.

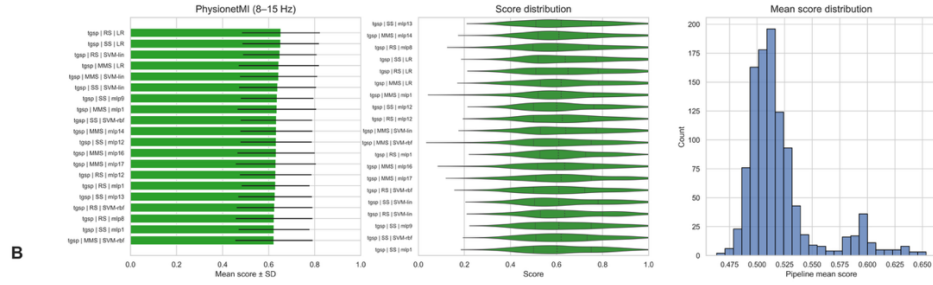
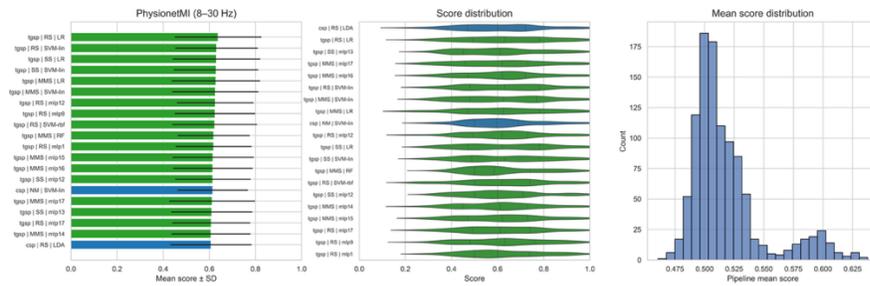
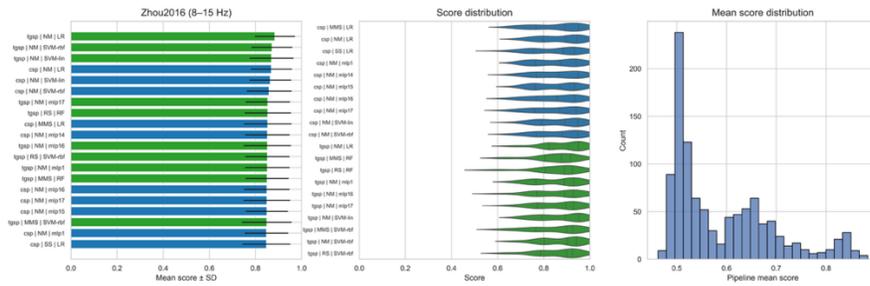
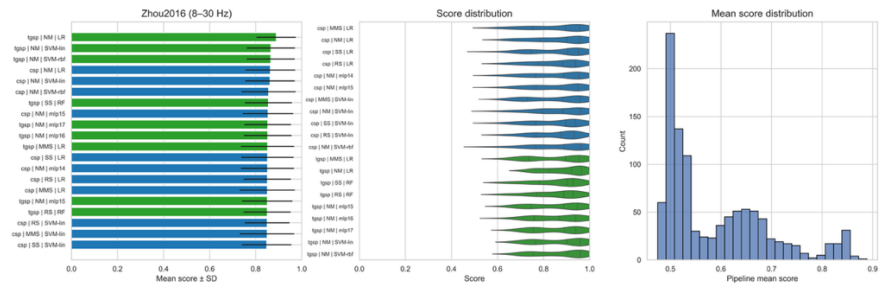

**Figure 1.** *Summary of classification pipeline performance for motor imagery EEG across two datasets (PhysionetMI and Zhou2016) and two frequency bands (8-15 Hz and 8-30 Hz). Each panel (A-D) summarizes three complementary views of pipeline performance: (left) the mean classification accuracy (±SD) of the top 20 pipelines, expressed using compact labels (feature extractor | scaler | classifier) to enhance readability (e.g., tgsp | RS | LR for Tangent Space + RobustScaler + Logistic Regression); (middle) the distribution of subject-wise scores for these top-performing pipelines; and (right) the global distribution of mean scores across all pipelines evaluated. Pipelines are additionally color-coded by family to facilitate interpretation (e.g., CSP in blue, Tangent-Space methods in green). Panels correspond to the four evaluation settings: (A) PhysionetMI 8-15 Hz, (B) PhysionetMI 8-30 Hz, (C) Zhou2016 8-15 Hz, and (D) Zhou2016 8-30 Hz.*

**Table 1.** *Mean classification accuracy of EEG pipelines (logistic regression), averaged across all subjects in PhysionetMI, Zhou2016, and Cho2017. Results are grouped by frequency band and preprocessing method.*

| Pipeline | Frequency Bands | Scores |
| --- | --- | --- |
| cov_tgsp+robustscaler+logistic_regression | 8-30 | 0.687612 |
| cov_tgsp+robustscaler+logistic_regression | 8-15 | 0.685674 |
| csp+standardscaler+logistic_regression | 8-30 | 0.655772 |
| csp+standardscaler+logistic_regression | 8-15 | 0.646789 |
| con_instantaneous_tgsp+StandardScaler+logistic_regression | 8-15 | 0.605672 |
| con_instantaneous_tgsp+StandardScaler+logistic_regression | 8-30 | 0.591630 |
| hfd+robustscaler+logistic_regression | 8-30 | 0.578524 |
| hjorth+standardscaler+logistic_regression | 8-30 | 0.576897 |
| svd_entropy+robustscaler+logistic_regression | 8-30 | 0.570882 |
| hjorth+standardscaler+logistic_regression | 8-15 | 0.556872 |
| svd_entropy+robustscaler+logistic_regression | 8-15 | 0.555421 |
| hfd+robustscaler+logistic_regression | 8-15 | 0.532427 |

Figure 2 shows the distribution of scores for the different pipelines (logistic regression) for both frequency bands confirming this trend. Cov-tgsp performs consistently well across both frequency bands, with a relatively high median score compared to other methods. The interquartile range (IQR) is smaller for the 8-15 Hz band than for the 8-30 Hz band, indicating that the scores for the 8-15 Hz band are more tightly clustered around the median. Both frequency bands show a high maximum score, reaching close to 1. CSP also shows a strong performance across both frequency bands, with a median score slightly lower than cov-tgsp. The IQR for CSP is wider for both frequency bands, suggesting more variability in its performance. However, the overall distribution is similar between 8-15 Hz and 8-30 Hz bands. In contrast, pipelines using non-linear features, such as hfd and svd_entropy, show increased variability with a moderate IQR, few outliers and lower scores overall, reflecting their poorer performance in classifying EEG signals in this study.

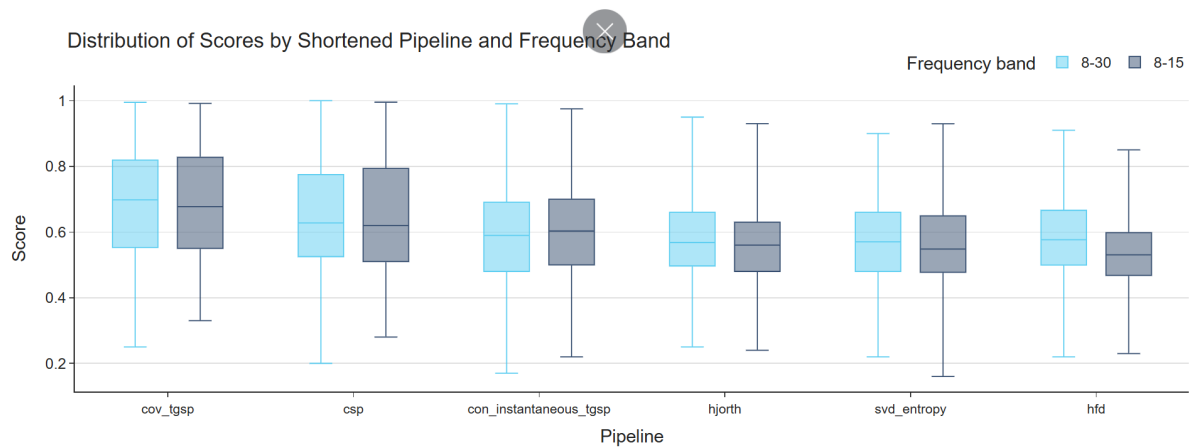

**Figure 2.** *Distribution of scores for various pipelines (logistic regression) across two different frequency bands: (8-15 Hz (in grey) and 8-30 Hz (in blue). The evaluation was made by combining the performance scores across all datasets and subjects (PhysionetMI, Zhou2016, Cho2017).*

**Inter-Dataset Benchmark**

Although cov_tgsp and CSP consistently lead in performance (see Table 2 & Figure 3), notable differences emerge across datasets. A dataset-wise analysis shows that, while CSP and cov_tgsp rank among the top performers in all cases, their effectiveness is noticeably lower on the PhysionetMI dataset compared to Cho2017 and Zhou2016. This discrepancy may be attributed to variations in experimental protocols, sampling rates, number of electrodes, and subject counts across the datasets.

Our findings highlight that the effectiveness of feature extraction methods in motor imagery EEG decoding is strongly dataset-dependent. For instance, on Zhou2016, the combination of CSP with StandardScaler and logistic regression achieved the best score (0.87), closely followed by cov_tgsp with RobustScaler and logistic regression (0.84). On Cho2017, cov_tgsp outperformed CSP, reaching a top score of 0.74 versus 0.68.

Table 2. *Mean classification accuracy of top pipelines (logistic regression), by dataset and frequency bands. Scores are averaged across frequency bands and subjects.*

| Pipeline | Bands | Dataset | Score |
| --- | --- | --- | --- |
| csp+standardscaler+logistic_regression | 8-15 | Zhou2016 | 0,87 |
| csp+standardscaler+logistic_regression | 8-30 | Zhou2016 | 0,87 |
| cov_tgsp+robustscaler+logistic_regression | 8-15 | Zhou2016 | 0,84 |
| cov_tgsp+robustscaler+logistic_regression | 8-30 | Zhou2016 | 0,84 |
| con_instantaneous_tgsp+StandardScaler()+logistic_regression | 8-15 | Zhou2016 | 0,75 |
| cov_tgsp+robustscaler+logistic_regression | 8-30 | Cho2017 | 0,75 |
| cov_tgsp+robustscaler+logistic_regression | 8-15 | Cho2017 | 0,73 |
| con_instantaneous_tgsp+StandardScaler()+logistic_regression | 8-30 | Zhou2016 | 0,72 |
| hjorth+standardscaler+logistic_regression | 8-15 | Zhou2016 | 0,71 |
| hfd+robustscaler+logistic_regression | 8-30 | Zhou2016 | 0,70 |
| con_instantaneous_tgsp+StandardScaler()+logistic_regression | 8-30 | Cho2017 | 0,70 |
| csp+standardscaler+logistic_regression | 8-15 | Cho2017 | 0,69 |
| svd_entropy+robustscaler+logistic_regression | 8-30 | Zhou2016 | 0,68 |
| svd_entropy+robustscaler+logistic_regression | 8-15 | Zhou2016 | 0,68 |
| con_instantaneous_tgsp+StandardScaler()+logistic_regression | 8-15 | Cho2017 | 0,67 |
| csp+standardscaler+logistic_regression | 8-30 | Cho2017 | 0,67 |
| cov_tgsp+robustscaler+logistic_regression | 8-15 | PhysionetMotorImagery | 0,64 |
| cov_tgsp+robustscaler+logistic_regression | 8-30 | PhysionetMotorImagery | 0,64 |
| hjorth+standardscaler+logistic_regression | 8-30 | Zhou2016 | 0,64 |
| hfd+robustscaler+logistic_regression | 8-15 | Zhou2016 | 0,63 |
| csp+standardscaler+logistic_regression | 8-30 | PhysionetMotorImagery | 0,63 |
| hfd+robustscaler+logistic_regression | 8-30 | Cho2017 | 0,61 |
| csp+standardscaler+logistic_regression | 8-15 | PhysionetMotorImagery | 0,60 |
| svd_entropy+robustscaler+logistic_regression | 8-30 | Cho2017 | 0,60 |
| hjorth+standardscaler+logistic_regression | 8-30 | Cho2017 | 0,60 |
| svd_entropy+robustscaler+logistic_regression | 8-15 | Cho2017 | 0,57 |
| hjorth+standardscaler+logistic_regression | 8-15 | Cho2017 | 0,57 |
| hjorth+standardscaler+logistic_regression | 8-30 | PhysionetMotorImagery | 0,56 |
| con_instantaneous_tgsp+StandardScaler()+logistic_regression | 8-15 | PhysionetMotorImagery | 0,56 |
| hfd+robustscaler+logistic_regression | 8-30 | PhysionetMotorImagery | 0,55 |
| svd_entropy+robustscaler+logistic_regression | 8-30 | PhysionetMotorImagery | 0,54 |
| svd_entropy+robustscaler+logistic_regression | 8-15 | PhysionetMotorImagery | 0,53 |
| hjorth+standardscaler+logistic_regression | 8-15 | PhysionetMotorImagery | 0,53 |
| hfd+robustscaler+logistic_regression | 8-15 | PhysionetMotorImagery | 0,53 |
| con_instantaneous_tgsp+StandardScaler()+logistic_regression | 8-30 | PhysionetMotorImagery | 0,53 |
| hfd+robustscaler+logistic_regression | 8-15 | Cho2017 | 0,52 |

In contrast, methods based on instantaneous functional connectivity and non-linear features showed lower performance across all datasets. For example, svd_entropy combined with RobustScaler and logistic regression yielded scores of 0.68 on Zhou2016, 0.59 on Cho2017, and 0.54 on PhysionetMI. On PhysionetMI specifically, overall performance was weaker, with the best pipeline (cov_tgsp) reaching only 0.64, and CSP averaging 0.62 across frequency bands. These results suggest that although non-linear and connectivity-based features capture complex dynamics of EEG signals, they may not provide superior discriminative power for motor imagery classification in this context.

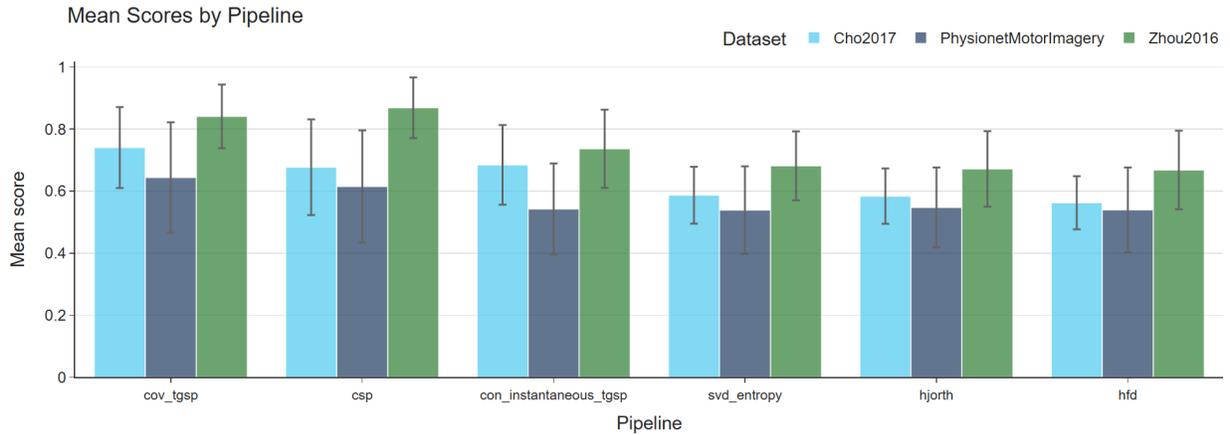

**Figure 3.** *Mean Scores by Pipeline (Logistic Regression) and Dataset (Averaged Across Both Frequency Bands)*

Figure 4 reveals a noticeable dispersion of scores across all pipelines and datasets, with the effect being particularly pronounced for PhysionetMI. Interestingly, this variability appears largely independent of the pipeline used, suggesting a high degree of EEG signal heterogeneity within this dataset. Such variability could stem from specific characteristics of PhysionetMI, such as its protocol, number of sessions, or experimental conditions, that complicate classification and reduce consistency across methods.

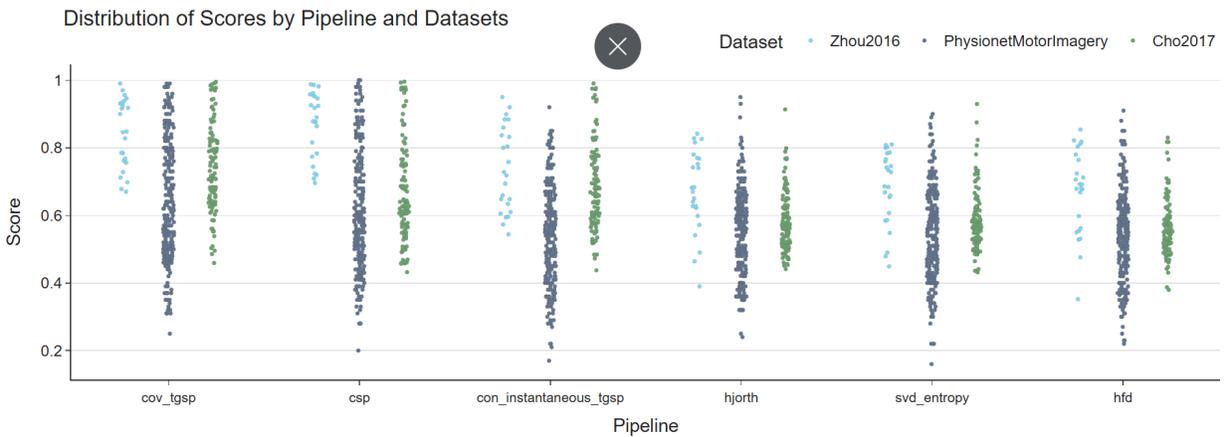

**Figure 4.** *Distribution of subject-level scores for each pipeline and dataset, averaged across frequency bands. In contrast, Zhou2016 and Cho2017 show tighter score distributions, indicating greater stability and more homogeneous signal properties. Among the pipelines, cov_tgsp and CSP exhibit the least dispersion, especially in Zhou2016, highlighting their robustness and consistent ability to discriminate motor imagery signals under more controlled conditions.*

**Intra-Subject Analysis**

Subject-level analysis reinforces the overall superiority of CSP and cov_tgsp across datasets, as illustrated in Figures 5 and 6 for Cho2017 and Zhou2016. However, this pattern becomes less consistent on the PhysionetMI dataset (Figure 7), where these pipelines do not systematically outperform others. In several cases, alternative methods achieve the highest scores: for instance, Hjorth parameters outperform CSP and cov_tgsp for participants #57 and #59, while the instantaneous connectivity approach combined with TGSP leads on participants #17 and #73.

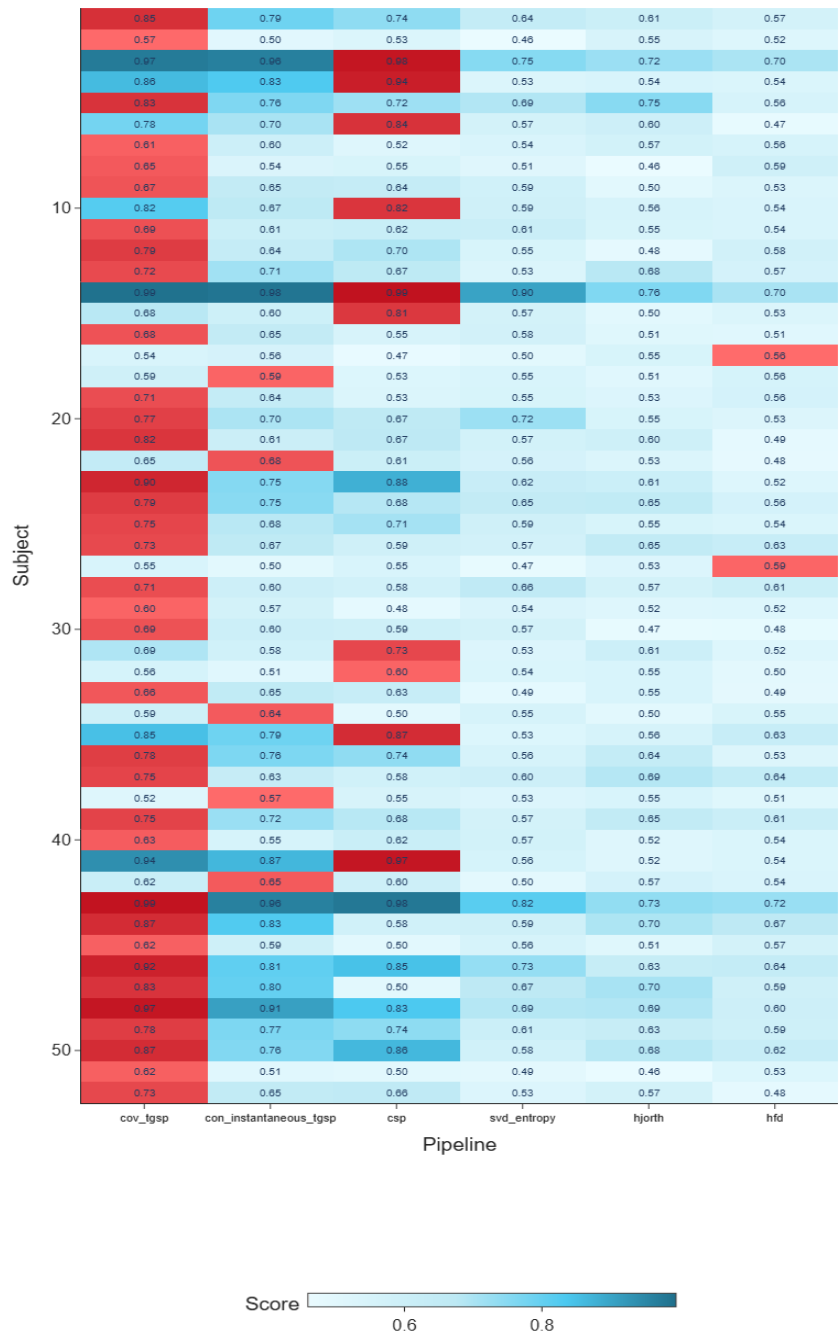

**Figure 5.** *Best Scores per Subject by Pipeline (in red) - Dataset Cho2017 (Averaged Across Both Frequency Bands)*

These results suggest that the specific characteristics of the PhysionetMI dataset (such as increased inter-subject variability, protocol complexity, or signal heterogeneity) may reduce the relative advantage of standard spatial methods. In this context, pipelines leveraging local temporal dynamics or connectivity features may prove more effective for certain individuals, highlighting the importance of personalized pipeline selection in EEG decoding.

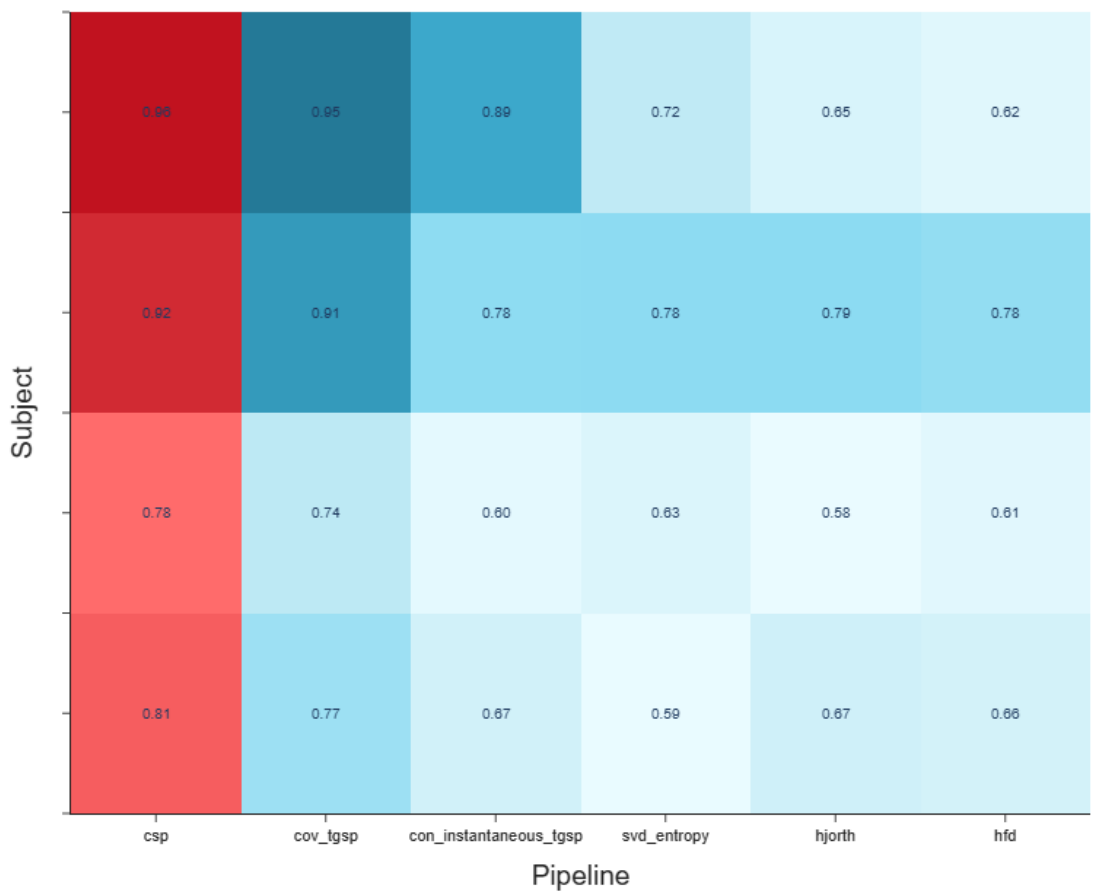

**Figure 6.** *Best Scores per Subject by Pipeline (in red) - Zhou2016 dataset (Averaged Across Both Frequency Bands)*

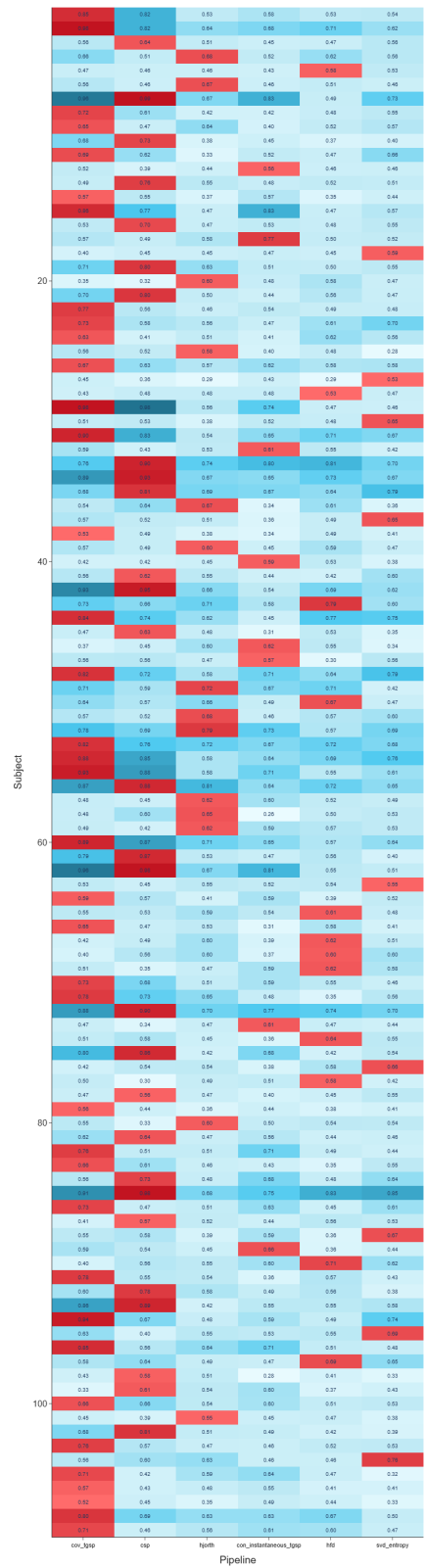

**Figure 7.** *Best Scores per Subject by Pipeline (in red) - Dataset PhysionetMI (Averaged Across Both Frequency Bands)*

**DISCUSSION**

Our findings reveal substantial variability in the performance of EEG decoding pipelines across datasets, underscoring the necessity of tailoring analytical approaches to individual participants for optimized BCI performance. Spatial filtering techniques, particularly covariance tangent space projection[49] and common spatial patterns[28], consistently delivered robust results on more homogeneous datasets such as Cho2017[14] and Zhou2016[39]. However, their effectiveness declined markedly when applied to the greater heterogeneity of the PhysionetMI dataset[57].

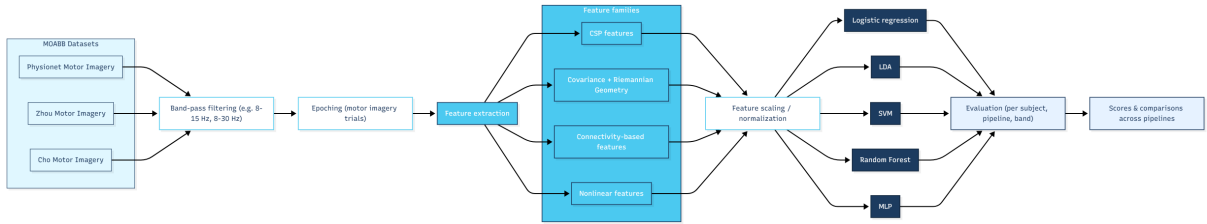

**Figure 8.** *Overview of the benchmarking pipeline applied to three publicly available motor imagery EEG datasets (PhysionetMI, Zhou2016, Cho2017) within the MOABB framework. EEG signals are band-pass filtered (8-15 Hz, 8-30 Hz), epoched, and transformed using multiple feature-extraction families, including CSP, Riemannian covariance-based methods, connectivity features, and nonlinear descriptors. Features are then scaled and classified using several classical machine-learning models. All pipelines are evaluated at the subject level and across frequency bands using a standardized within-session protocol, enabling systematic comparison of decoding performance and inter-individual variability.*

This suggests that generalized pipelines (see Figure 8), though effective under controlled conditions, may falter significantly when confronted with the diverse physiological and neurophysiological profiles encountered in real-world scenarios[19,45]. This underscores the importance of considering the unique neurophysiological profiles of each participant. It also highlights the necessity of customizing classification models for real-life conditions to better accommodate individual neurophysiological idiosyncrasies. For instance, subject-specific functional connectivity patterns have been shown to exceed task-induced variability[70]. Additionally, inter-subject EEG variability remains markedly greater than intra-subject variability, suggesting that such variability is rooted in stable individual traits rather than momentary task effects[16].

An important challenge in current BCI research, as well as the use of such systems in real life settings, is the limited understanding of the neural mechanisms underlying individual differences observed in EEG signals[3,13,16]. While spatial and nonlinear feature extraction methods have proven highly effective in enhancing EEG-based classification performance, particularly for tasks such as motor imagery, emotion recognition, and mental workload detection, their success remains limited by an inability to fully capture the neurophysiological idiosyncrasies of individual subjects. EEG signals are both non-stationary and subject to strong individual differences, driven by factors like anatomical traits, genetic factors, electrode positioning, and fluctuating neural states, leading to shifts in input distributions and undermining cross-participant generalization[71].

These methods often struggle to disentangle genuine neural signals from physiological noise, especially in heterogeneous populations leading to poor generalization of models trained form one participant to another. A recent study on cross-subject EEG-based emotion recognition highlights that anatomical, cognitive, perceptual, and hardware differences across individuals significantly degrade model generalization[72]. However, studies increasingly advocate the integration of complementary biosignals such as electromyography (EMG), electrooculography (EOG), and heart rate variability (HRV) to better isolate neural activity and improve the robustness of decoding pipelines in real-world BCI applications[20,73]. Moreover, EEG's inherently limited spatial resolution and susceptibility to volume conduction effects constrain its ability to differentiate fine-grained neural patterns, reinforcing the necessity for personalized multimodal pipelines that leverage complementary data streams[19,66].

The marked variability observed, particularly within the PhysionetMI dataset, likely arises from its inherent complexities such as diverse experimental conditions, subject-specific neurophysiological traits, and differences in signal quality[40,50]. Previous research underscores that EEG signal characteristics can fluctuate significantly due to inter- and intra-subject variability factors, including fatigue, attention shifts, and session-to-session differences[19,27].

Our work reinforces these observations by illustrating how spatial and nonlinear methods variably capture relevant information depending on individual participant characteristics and dataset-specific conditions. The limitations of EEG decoding pipelines further extend to their ability to handle non-stationarities in EEG signals: a persistent problem that hampers the long-term usability of BCI systems. EEG signals are inherently non-stationary due to changes in cognitive state, fatigue, attention levels, and learning effects over time[63]. Traditional approaches, such as CSP and TGSP, typically assume stationarity, resulting in degraded performance over longer periods or across different experimental sessions[47].

Developing EEG data processing and classification pipelines capable of continuously adapting to temporal variations, potentially through reinforcement learning or adaptive online algorithms, remains a key direction for future research to enhance real-world viability[19,73].

An important aspect of our findings is the substantial improvement that individualized pipeline selection can offer over a one-size-fits-all strategy. This is particularly evident in the subject-level analysis of the PhysionetMI dataset, where alternative methods such as Hjorth parameters[15] and instantaneous connectivity combined with tangent space projection outperformed CSP and cov-TGSP for several subjects. These results underscore the necessity for adaptive or hybrid methodologies, as suggested in recent literature[58,63]. Such approaches, which dynamically adjust their parameters or even feature extraction techniques according to subject-specific EEG dynamics, could significantly mitigate the challenges posed by the observed variability.

Our electrode-wise nonlinear feature analysis indicated that localized temporal dynamics are also crucial for decoding performance. Methods capturing local EEG complexity, such as Higuchi fractal dimension, singular value decomposition entropy, and Hjorth parameters, provide complementary information to traditional spatial filters, potentially reflecting subtle neuronal interactions or cognitive engagement states not captured by purely spatial approaches[38,48]. A promising direction for future research involves integrating these nonlinear features within spatial filtering frameworks or ensemble methods (e.g., stacking or boosting) to enhance the robustness and generalizability of EEG decoding algorithms across varying experimental contexts and subject-specific neurophysiological profiles[47,65].

Our results also revisit and offer a new perspective on the longstanding challenge of BCI illiteracy, traditionally described as the inability of some individuals to effectively operate a BCI despite training[17,18]. Rather than reflecting intrinsic user limitations, our findings suggest that illiteracy often stems from mismatches between the neurophysiological characteristics of individual users and the decoding algorithms employed. Consequently, personalized EEG decoding pipelines could substantially reduce instances of BCI illiteracy by accounting for individual neurophysiological idiosyncrasies[19,20].

The MOABB benchmark, the largest reproducibility study in EEG-based BCI research (Aristimunha 2023), evaluated 30 pipelines encompassing raw, Riemannian, and deep learning approaches across 36 publicly available datasets covering motor imagery, P300, and SSVEP paradigms. Its results indicate that Riemannian geometry-based methods, which exploit the intrinsic manifold structure of covariance matrices, consistently outperform both raw-signal approaches and deep learning models in terms of robustness and overall accuracy[31,58,62,65,74].

While deep learning methods hold great promise, their practical success in EEG decoding remains contingent upon extensive, well-curated training datasets and meticulous hyperparameter tuning[73]. Our findings further underscore the importance of developing subject-specific models to effectively address the variability inherent in EEG signals for real-world BCI applications[31,61]. Consequently, future efforts should explore hybrid strategies that combine classical spatial and nonlinear methods with deep learning models optimized for individual variability.

One particularly promising direction is ensemble learning: bagging, boosting, or stacking could integrate the complementary strengths of linear spatial filters and nonlinear dynamics, yielding decoders that remain robust across the full spectrum of inter- and intra-subject variability[31,75]. To validate and stabilize such approaches, future work should replicate these benchmarks across additional paradigms (SSVEP, P300, CVEP) and on invasive intracortical recordings, where signal-to-noise ratios and spatial specificity differ markedly[59,61]. Parallel efforts should also compare advanced deep architectures, such as CNNs, LSTMs, and convolutional transformers, within a unified evaluation framework, leveraging libraries like Braindecode[19,64].

The emerging field of quantum machine learning[75,76] presents exciting opportunities for managing the inherent noise and complexity of EEG data. Despite the promise of quantum machine learning approaches such as quantum support vector machines (QSVM), their practical application to EEG decoding still faces significant challenges[77–79]. Quantum methods theoretically exploit high-dimensional Hilbert spaces to improve classification performance; however, recent studies highlight critical issues, including unstable gradients during training and difficulties in scaling quantum algorithms to effectively handle the large dimensionality of EEG data[80,81]. These limitations suggest that, while quantum machine learning may enhance robustness and discrimination capability for smaller-scale EEG tasks, substantial advances in quantum computing infrastructure and algorithms are required before it can reliably address the complexity and dimensionality of large-scale EEG decoding.

A major obstacle, widely known as the "barren plateau" problem[82], has emerged as a key limitation in training quantum neural networks such as variational quantum circuits. Barren plateaus are characterized by vanishing gradients, which make optimization exceedingly difficult or practically infeasible for circuits of even moderate depth or complexity[83,84]. Recent studies have shown that the exponential dimensionality of Hilbert spaces, although theoretically advantageous, can exacerbate barren plateaus due to concentration-of-measure phenomena. This means that increasing the complexity or number of features in quantum models (as might be required for high-dimensional EEG data) can rapidly degrade learning efficiency, causing the quantum approach to perform worse than classical counterparts.

Consequently, current research advocates cautious optimism toward quantum-inspired EEG decoding until effective strategies such as specialized initialization schemes, shallow architectures, or hybrid classical-quantum methods are fully developed and validated in practice. Nevertheless, quantum machine learning holds considerable potential as a transformative approach for EEG decoding. Recent breakthroughs have demonstrated that carefully designed quantum circuits with appropriate initialization schemes and ansatz structures can maintain stable gradients, paving the way for scalable and robust EEG classifiers[85,86]. Moreover, inherent quantum properties such as entanglement and superposition offer unique avenues for capturing and processing complex EEG signal dynamics that classical models struggle to represent efficiently.

Therefore, despite current limitations, quantum approaches, when combined with strategic algorithmic advances, could ultimately deliver substantial improvements in decoding and classificaiton accuracy of EEG signals, generalization, and resilience to EEG's intrinsic variability, marking a major step forward in the development of personalized and adaptive BCIs.

Our ongoing preliminary studies within the quantum computing framework aim to rigorously test these theoretical advantages, providing empirical insights into whether quantum-inspired methodologies can concretely enhance EEG decoding performance.

Another crucial limitation observed across current EEG decoding methodologies is their reliance on labeled data, which can be prohibitively expensive and time-consuming to obtain at scale. Recent literature emphasizes the need for unsupervised or semi-supervised learning techniques capable of leveraging unlabeled EEG data, thereby reducing dependence on extensive labeled datasets[20]. Furthermore, transfer learning approaches, capable of transferring neural patterns learned from one subject or dataset to another, remain underexplored despite their potential to substantially improve model generalization and mitigate the effects of inter-subject variability[47,73]. Future research should prioritize the development and benchmarking of transfer learning pipelines to assess their viability in overcoming dataset-specific limitations observed in EEG decoding.
In summary, our benchmark supports the adoption of individualized EEG pipelines to effectively address inter- and intra-subject variability. The integration of adaptive strategies, nonlinear dynamics, ensemble learning techniques, and quantum-inspired methodologies collectively represents the frontier of personalized and clinically relevant BCI technologies, essential for transitioning EEG-based systems from controlled experimental settings to reliable, real-world applications.

Finally, ethical considerations surrounding personalized EEG decoding require careful attention[87]. As BCI technologies move closer to clinical and everyday deployment, ensuring the privacy, security, and ethical management of neural data becomes paramount. Thus, alongside technical innovation, future research should proactively address ethical implications, fostering responsible development and reinforcing user trust in EEG-based BCIs.

**CONCLUSION**

Our large-scale benchmark reaffirms that spatial methods, particularly covariance tangent-space projection and Common Spatial Patterns, remain the gold standard for EEG motor imagery decoding under controlled conditions. Nevertheless, their efficacy significantly deteriorates as inter- and intra-subject variability increases.

By exhaustively evaluating each pipeline at the subject level, we conclusively demonstrate that no universal configuration optimally suits every individual. Instead, our findings underscore that the most promising path forward lies in adaptive methodologies that dynamically align with each user's unique neurophysiological signature. Additionally, certain nonlinear features, such as Hjorth parameters, Higuchi fractal dimension, and singular value decomposition entropy, demonstrated substantial potential for specific individuals. This observation suggests that capturing richer neural dynamics can significantly enhance personalized decoding and indicates that "BCI illiteracy" may be more accurately attributed to mismatches between decoding algorithms and user-specific neurophysiological characteristics rather than inherent user limitations.

Future research should prioritize the exploration of sophisticated ensemble learning frameworks that integrate linear spatial filters with nonlinear dynamics, effectively harnessing their complementary strengths. Extending this benchmark to encompass broader paradigms such as SSVEP, P300, CVEP, and invasive intracortical recordings will further test the generalizability of our conclusions. Additionally, rigorous evaluation of state-of-the-art deep-learning architectures (CNNs, LSTMs, Transformers) alongside emerging quantum-inspired models, including quantum support-vector machines and hybrid quantum neural networks, represents an essential frontier. These quantum methodologies, leveraging the uniquely high-dimensional properties of Hilbert spaces, may offer unprecedented robustness against the intrinsic noise and complexity of EEG signals. Another critical direction involves improving cross-subject generalization through transfer learning and domain adaptation approaches, which have demonstrated strong potential to enhance model robustness and scalability across heterogeneous user populations[59,88]. Equally important is the need to strengthen reproducibility through open-access benchmarking initiatives such as MOABB, which enable standardized, transparent comparison of EEG decoding algorithms[59].

Finally, as personalized EEG decoding and adaptive BCIs move closer to clinical and real-world deployment, ensuring the ethical, private, and secure management of neural data must remain a central concern[89,90]. Responsible innovation and user-centered design will be essential to fostering trust and ensuring that advances in EEG decoding translate into safe, inclusive, and effective technologies.
Collectively, these advancements promise to evolve our understanding of what works, for whom, and under what conditions into practical, reliable, and clinically viable BCIs capable of robustly adapting to diverse user populations and real-world scenarios.

**CONFLICT OF INTEREST**

Olivier Oullier and Paul Barbaste are co-founders of Inclusive Brains, which has an active research partnership with IBM. Xavier Vasques is an employee of IBM. The collaboration between Inclusive Brains and IBM involves joint research on artificial intelligence, neurotechnology and quantum informatics but did not influence the design, data collection, analysis, or interpretation of results in this study. The remaining aspects of the work were conducted independently for academic purposes. No additional financial or non-financial conflicts of interest are declared by the authors.